# E-polis: A serious game for the gamification of sociological surveys

Alexandros Gazis*
*Democritus University of Thrace*
*School of Engineering*
*Department of Electrical and Computer Engineering*
Xanthi, Greece
agazis@ee.duth.gr
https://orcid.org/0000-0001-7146-9170

Eleftheria Katsiri
*Democritus University of Thrace*
*School of Engineering*
*Department of Electrical and Computer Engineering*
Xanthi, Greece
https://orcid.org/0000-0002-6215-2574

*Abstract*— E-polis is a multi-platform serious game that gamifies a sociological survey for studying young people's opinions regarding their ideal society. The gameplay is based on a user navigating through a digital city, experiencing the changes inflicted, triggered by responses to social and pedagogical surveys, known as "dilemmas". The game integrates elements of adventure, exploration, and simulation. Unity was the selected game engine used for the development of the game, while a middleware component was also developed to gather and process the users' data. At the end of each game, users are presented with a blueprint of the city they navigated to showcase how their choices influenced its development. This motivates them to reflect on their answers and validate them. The game can be used to collect data on a variety of topics, such as social justice, and economic development, or to promote civic engagement and encourage young people to think critically about the world around them.

*Keywords*— Training, Virtual Reality, User Interface, Serious Game, Serious Digital Game Middleware Architectures, Education Serious Games, Educational Art Games, Novel Serious Game, Game Development, Virtual Reality Digital Games

## I. Introduction

Based on Moore's law, [1], while the capabilities of computing resources processing are increasing, the overall cost of devices and their integrated circuit components for upgrading them are reduced, [2]. In other words, the devices used in our everyday lives, including smart cellphones, tablets, laptops, workstations, and personal computers, are increasingly powerful in terms of CPU utilization (available cores and threads) and system load (RAM and I/O operations speed), [3], [4]. As a result, during the past decades, there was a surge of interactive media that operate on smart low-power, and low-cost electronic devices, [5], [6]. In the context of our study, "interactive media" means webpages, user-generated content, television, advertising, blogs, and digital games, [7], [8], [9], [10].

Nowadays, due to the processing power and the wider usage of internet access, digital games using complex software applications and either 2D or 3D graphics are now available to a large audience, [11], [12]. Specifically, "digital games" are generally defined as computer platform applications that enable two (or more) players to interact either through a network or locally via a computer, [13]. Digital games can be used as a smart tool for entrainment or pedagogical purposes (i.e., to "deliver, support, and enhance teaching, learning, assessment, and evaluation"). Based on this theory, digital games can be categorized as serious or non-serious, based on their purpose, i.e., if they focus on education or pure entertainment, respectively, [14].

Based on the above, this article focuses on serious games (SG). Specifically, SGs are digital games that solely focus on serving an educational or training purpose. This term is not new, since early SGs can be found in the mid-90s, when the game "Brown Box" was produced to train soldiers on the use of radar sensory devices. Serious games can also be entertaining for players; however, their main goal is to provide a framework for players to develop their knowledge, showcase their threats, and promote social change and understanding. In detail, SGs have been developed in a vast number of different areas, such as the military, law, training, education, environmental issues, social issues, and healthcare, [15].

As such, our research paper aims to create a serious game that allows users to navigate a digital city and envision an ideal society. The game is designed to present social and pedagogic questions to players, which must be answered to cross the streets of a town/neighborhood. The purpose of the game is to simulate a virtual environment for representing democracy by giving choices (actions) to the players. The final game design is based on MIT's design principles, after studying several modern approaches and serious games.

The objectives of our research paper are twofold: technical and theoretical. From a technical perspective, we aim to develop a cloud computing middleware that fuses different software layers of the game into a single application, mainly focusing on data collection, aggregation, and question-answer processing. We also aim to store players' answers, using certain layers (components) of our middleware in text format (CSV files or JSON format), as well as using a local DB file (SQLite). From a theoretical perspective, we aim to design a new conceptual social framework to identify, select, and evaluate various quality attributes. Our serious game examines young people's thinking and actions, as well as how they envision an ideal community.

Specifically, in the following sections, firstly, we present the main aspects of our research paper on digital serious games. Secondly, we review the related work on the categories, definitions, and characteristics of serious games, as well as the platformer genre of digital games. Thirdly, we describe the technical and theoretical novelty of our approach, which combines elements of serious games and platformers to create an engaging and educational game. Fourthly, we explain the implementation details of our game, such as the game engine used, the Unity scenes created, and the data collection methods. Finally, we conclude with a summary of our contributions and future work.

## II. RELATED WORK

Firstly, we present the genres of digital games and expand on their definition, while presenting examples of serious and non-serious games. Then, we focus on serious games and present certain areas they are applied to. Finally, we focus on the Platformer type of game i.e., a mix of the above categories and the type of game we have developed and described in our article.

### A. Categories of Digital Serious Games

Digital serious games, similar to traditional serious games, can be categorized into strategy, real role-playing, massively multi-role playing, government, and general simulation genres. The most popular case examples for these categories are the following, [16], [17]:

- Strategy games: Each decision made by the player has consequences on the outcome of the game. In addition, the vast majority are the so-called 'war games'. In strategy games, the player can observe how virtual worlds operate. The most popular games in this category are Warcraft, Crusader Kings, StarCraft, Total War, Sid Meier's Civilization, Age of Mythology, and Heroes of Might.

- Real Role-playing Games: These games are also called 'RPG' after the initials of the category. The player is not confined to strict strategy game sets actions and does not aim to observe the virtual world/online community to understand its whereabouts. On the contrary, the player focuses on navigating and choosing among available quests. Some of the best-known examples of this category are Diablo, Witcher, Pillars of Eternity, Elder Scrolls, Fallout, Dark Souls, and Mass Effect.

- Massively Multi Role Playing Games: These games consist of multiple virtual worlds that focus on the player's interaction and free will on the outcome. These digital worlds integrate players into small independent communities. Specifically, this is achieved through digital objects and avatar profiles which are digital entities representing each authenticated user. Finally, online worlds/communities usually operate throughout the day, regardless of country and time zone, and typically require a certain amount of fee or type of subscription (card) for the hours the user is logged in. The most well-known games in this category are World of Warcraft, Lord of the Rings Online, Active Worlds, Croquet Project, Project Wonderland, Second Life, EverQuest, and Guild Wars.

- Simulation games: This type focuses on simulating real-life scenarios or specific tasks usually defining and obeying the laws of nature and real-life case scenarios. The most well-known games in this category are The Sims, Sim City, Sim Ant, Sim Earth, Train Sim World, World of Warships, and Flight Simulator

- Government Simulation Games: This category has many similarities with simulation games, but specializes in topics. Analytically, subjects of interest are governmental, political, geopolitical legislation, and other internal political developments. Some of the best-known examples of this category are Tropico, Europa Universalis IV, World of Workshops, Government Simulator, Oligarchy, and Democracy.

### B. Serious Games

The main design principles of digital games development focusing on SG development are referred to in recent literature as "Game-Based Learning (GBL)". In GBL, the aim is to advance the players' education/learning experience, instead of entertaining them, [18]. In detail, SGs contribute to students applying factual knowledge, learning on demand, i.e., without a strict timetable, and coming into contact with new situations to gain new experiences and knowledge, [19], [20], [21], [22]. SGs can make the learning process more effective and efficient, as they assist in maintaining high levels of interest among participants and promoting the active involvement of users in problem-solving, [23], [24], [25], [26]. Thus, SGs are a particularly important training tool as they follow the so-called "situated learning model", [27], [28], [29]. Over the years, SGs continuously incorporate new features and are enriched with new game and graphical methods.

In Table I, we expand the work of, [30], [31], and present a taxonomy of the fields of study of serious games.

TABLE I. A TAXONOMY OF SERIOUS GAMES FIELDS OF STUDY, APPLICATIONS AND DESIGN PRINCIPLES

| Field of Study | Applications | Serious Game Design Principle |
|---|---|---|
| Pedagogy | Museum Spaces | Human-Computer Interaction (HCI) |
| Cognitive | History | History Presentation |
| Learning | Mathematics | Objectives |
| Psychiatry | Technology | Rules and Lists of Quests |
| Logic Choices | Social Studies | Controls and suggestions via Artificial Intelligence or Data processing |
| Perception | Linguistic Research | 2D/3D graphics |
| Ethical | Human values | AI ethics rules |
| Regulatory | Laws-Regulations | |

### C. Platformer Digital Games Genre

Platformer Digital Games emerged in the 1980s. They provide players with more freedom when it comes to movement, quest selection, and orientation/direction choices, typically in a specific and strictly predefined action environment, [32]. Platformer Digital Games are a mix of Real Role-playing Games, simulation, and strategy games. In particular, a Platform game is characterized by the frequent use of navigation options (e.g., walk, run, swim, and climb) to achieve a goal. The player needs to study the different terrains, surpass possible obstacles (thorns, volcanoes, steep rocks, etc.), and understand the general rules and logic of the game to accelerate, stop, jump, climb, and overall choose the right path to reach the finish line of each level (quest). As such, it is customary before the finish line that players are presented with multiple options to collect items to boost or upgrade their experience (e.g., coins, keys) as well as to face an opposing character, known as the "boss", [33]. Finally, it is worth mentioning that adventure games and puzzle games are similar in that they both require players to solve puzzles and navigate through a game world. In both genres, players use a set of predefined movements and actions to interact with the environment. As a result, both genres require players to make choices throughout gameplay and use problem-solving skills to overcome obstacles.

Some examples of these games are Space Panic and Donkey Kong. Similarly, other notable applications that contributed to the popularity of this category are Super Mario, Mario Brothers, Pitfall, Sonic the Hedgehog, and, more recently, Super Meat Boy Forever, Celeste, Limbo, and Alto's Odyssey.

III. THE RESEARCH PROBLEM

The digital game that was developed in the context of this project is classified as SG because it is not exclusively aimed at entertaining the user. Its design principles study the players' behavior and choices while training them on how their actions could shape our society. As a result, this SG can be described as a modern research tool that may be widely used by both the social sciences and education. Specifically, this game aims to understand the characteristics of young persons by utilizing both quantitative and qualitative design methods. To do that, we created a digital world where players are free to roam a digital city. They are presented with quests in the form of questions. Since this is a SG, there are no correct or incorrect answers. The players navigate through a city and when they answer all the available questions, they are redirected to a new scene where they see how these choices shaped the city. It is up to them to judge whether they participated in its advancement or deterioration.

In this manner, players reflect on their decisions and gain an in-depth understanding of critical socio-political issues of a democratic society. As a result, our game may serve as a research tool that, in addition to providing users with questions and data collection, will record their visions, decisions, and reactions to decision-making processes, forms of political participation, organization, and ultimately the functioning of society as a whole, [34].

IV. TECHNICAL NOVELTY

Given that serious games are tailor-made, and they intend to teach new methods and advance players' knowledge, they are novel, by design. The purpose of our study is to create an SG that allows users to navigate a digital city and envision an ideal society. Specifically, players are presented with social and pedagogic questions that must be answered to cross the streets. There is no option to skip or choose not to answer a given question.

Additionally, as this is an SG, no correct or false answers apply. We aim to simulate a virtual environment for representing democracy by giving choices (actions) to the players. The final game design is based on MIT's design principles, after studying several modern approaches and SGs [35], [36], [37], [38], [39], [40], [41], [42], thus, providing a game that covers all aspects of human-computer interaction.

As such, our game novelty is twofold:

1. **Technical**: Since Unity is an "all battery included" framework, our technical expertise was not focused on the development of the game, but rather on the software entities' interaction. We developed a middleware that fuses different software layers of the game into a single application, mainly focusing on data collection, aggregation, and question-answer processing. As such, we highlight the following aspects:

a. We chose to store players' answers in text format in CSV files or JSON format, as well as using a local DB file and specifically, using SQLite, [43], which is one of the most versatile and low-resourcing demanding DBs. All of these choices are not officially supported out of the box when installing Unity. Specifically, this DB as well as all other data formats schemas generated are not included in Unity's packages or applied modules

b. We created a middleware that similarly to [44], [45], categorizes each software entity based on its application module, behavior, frequency of instances, and prefabs

c. We defined preset rules regarding players' actions. Specifically, based on their answers, we redirect players to new scenes, where they witness the change based on their overall choices and answers to questions posed. To this end, the technical novelty lies both in the transition as well as the development of each player's decision that spans from early question-answers to a predefined set of behaviors. For example, apart from answers, we store how much time is needed to answer each question (thus, we detect if the question was intriguing or difficult to answer) as well as the player's exact location and orientation (thus, we detect the questions and quests which are of high interest).

2. **Theoretical**: We have designed a new conceptual social framework to identify, select, and evaluate various quality attributes. The game studies the expected social relations and management practice, based on the players' answers to preset questions. To this end, our SG examines young people's thinking and actions, as well as how they envision an ideal community, [46], [47].

V. IMPLEMENTATION

*A. Game Engine Used*

Since the game's purpose is educational, there was no need to use high computer resources or complex graphics and animations. As such, we studied the most widely used game engines in the industry (OpenSimulator, Godot, CryEngine, GameMaker) and chose the Unity and Unreal engines. Both of these tools are complex frameworks that provide the necessary tools to develop and export our game to a single deliverable execution file for players to interact. Upon careful consideration, we chose Unity. Although it provides fewer options in terms of scaling the application, its use allows us to focus on the game's development, irrespective of the final platform or operating system we

deploy - thus, its pros outweigh the cons. In detail, during the early stages, we had not chosen which platform (i.e., IOS, Android, and Windows) was suitable for players to interact with the game. Thus, Unity enabled us to develop our project without the need to consider software dependency issues, package recompilations, and common pitfalls that exist when creating a cross-platform application, [48]. For the development of this game, we used Unity v.2020.3.24f1 and Unity's private license.

Lastly, the hierarchy of our game objects is the following (Figure 1):

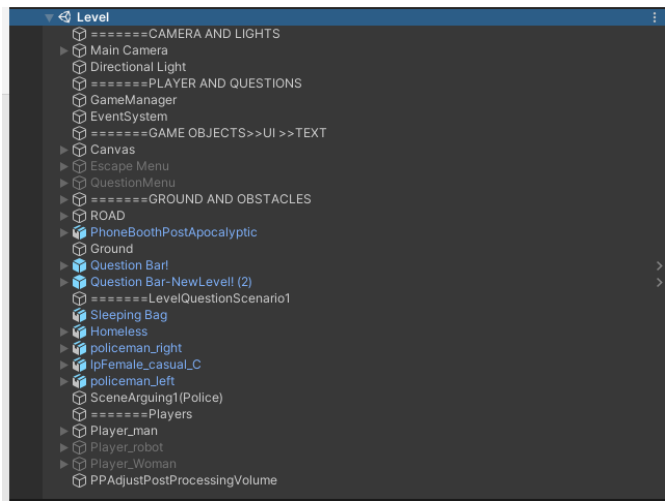

Fig. 1. Unity's hierarchy view of our serious digital game

It is worth noticing that, although the Hierarchy option classifies the game objects, it mainly affects the Unity engine, as these objects are the link to the navigation windows (Inspector) where we attach new properties.

*B. Unity's Scenes*

In Unity, each scene represents the level where the player interacts.

The first level of our game is the Introduction menu, where the player is presented with the game's "Main Menu", i.e., options to start the game, select the player's profile, configure the music volume level, and exit the game.

The second scene of our game represents an actual place in Greece, located in Athens, Monastiraki area, called "Plateia Agias Eirinis", which was recreated digitally for players to navigate (Figure 2).

Lastly, based on the players' choices of the player, when all questions are answered, they are redirected to the final scene which provides a view from above i.e., a blueprint of the city after the player's choices are presented, thus giving players the opportunity to understand the overall impact of their answers.

In the next sections, we present the scenes of our game and an indicative question posed to users. It is emphasized that the graphic elements, color schemes, and overall visualizations presented have been created exclusively by the developers of the application and are therefore subject to change in future versions of the game.

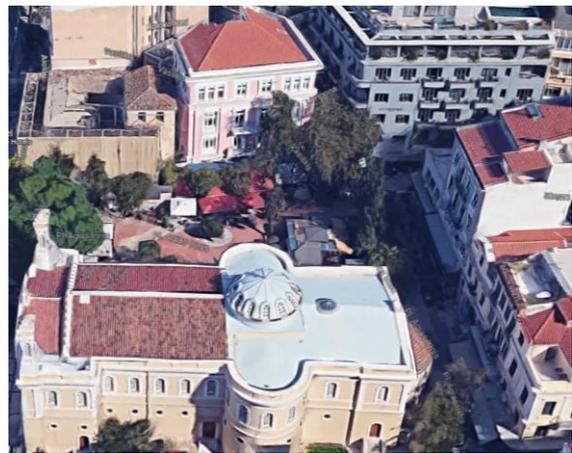

Fig. 2. Outside view of our actual case study "Plateia Agias Eirinis", Athens, Monastiraki area, Greece

*1) Unity Scene 1 – Introduction Menu*

Initially, when the game starts, the player is presented with the following scene (Figure 3):

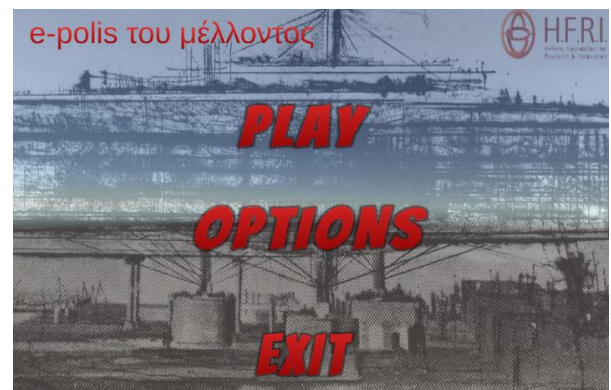

Fig. 3. Introduction Menu to our serious digital game

As such, when players select the first option (i.e., "PLAY"), they can choose their character (Figure 4). Once a character is chosen, they can start playing the game.

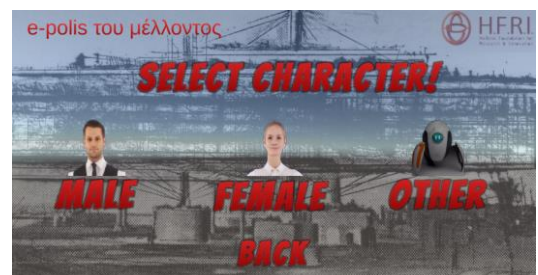

Fig. 4. Introduction Menu, "START" Menu

Similarly, when the player selects the second option (i.e., "OPTIONS"), a new menu is displayed which allows users to configure the game's volume (Figure 5).

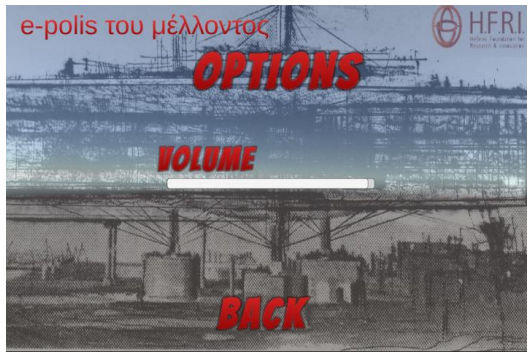

Fig. 5. Introduction Menu, "OPTIONS" Menu

*2) Unity Scene 2 – Real World View*

Below, we present illustrations of Scene 1 from different angles to understand its structure and dimensions (Figure 6):

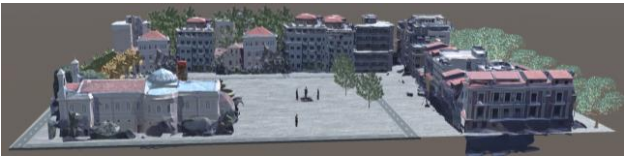

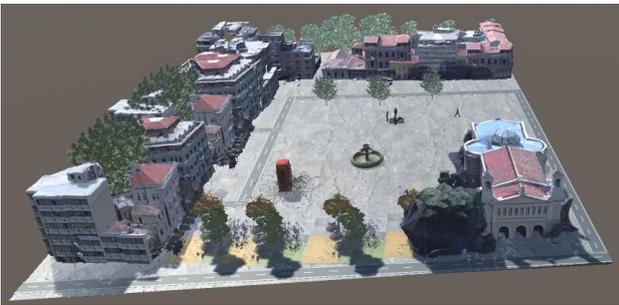

Fig. 6. Actual gameplay's 3D depiction of the real-world view of our scene

In addition, it is emphasized that players can change the angle of the camera and therefore the way the game is displayed, by rotating the mouse or zooming in and out of the graphical representation, through the use of their mouse wheel (fluctuating the angle of view).

The purpose of this scene is to pose players with various dilemmas. In particular, when starting the game, the player can navigate the world using their keyboard as well as change their perspective (angle of view as well as fluctuation of the field of view) using their computer mouse. In addition, as we previously pointed out, the purpose of the game and navigation is to present the player with various (social) dilemmas.

For example, while navigating near the starting point, the player is presented with 3 police officers who shout at a homeless man, as shown in Figure 7. When the player gets close enough to examine this incident, a dilemma appears with available answers – actions: "How would you react? A. Confront the police officers and stand up for the man, B. Not confront the police officers, but video-record their actions, C. Congratulate the police, D. Leave; there is nothing to do here".

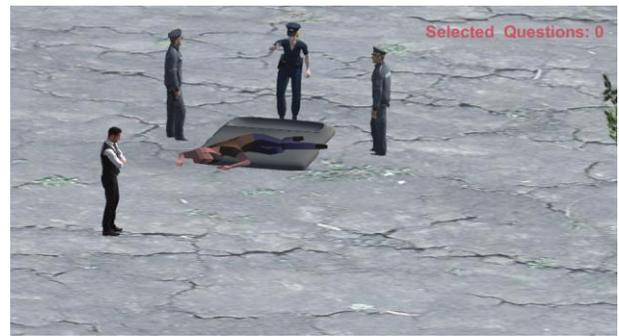

Fig. 7. Actual gameplay's 3D depiction of an incident occurring in the digital world of the serious digital game

Also, the user's answers are stored by our computer system (both the question and the answers given) to be used for the analysis and processing of the survey data, taking into consideration the socio-political extensions of the research project. When the players answer said dilemmas, they are required to visit the telephone booth strategically placed in a specific path which must be followed by all players. The telephone booth signals the end of this quest (Figure 8):

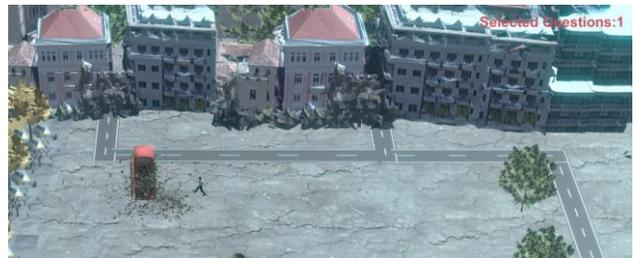

Fig. 8. Phone booth that signals the end of this quest and the transition to the next

In addition, it is worth mentioning that the number of the user's answers is also displayed during the game posting in the right corner of the user's screen (see Figure 7 and Figure 8 the top right corner where red letters showcase the number of answers). Finally, based on the logic and operating principles of the game, the player is an entity with physical dimensions, thus they may not "interact" with a building i.e., climb walls, enter a building, or jump on their roofs.

*3) Unity Scene 3 – View From Above*

Below, we present illustrations of Scene 3 from different angles to understand its structure and dimensions (Figure 9):

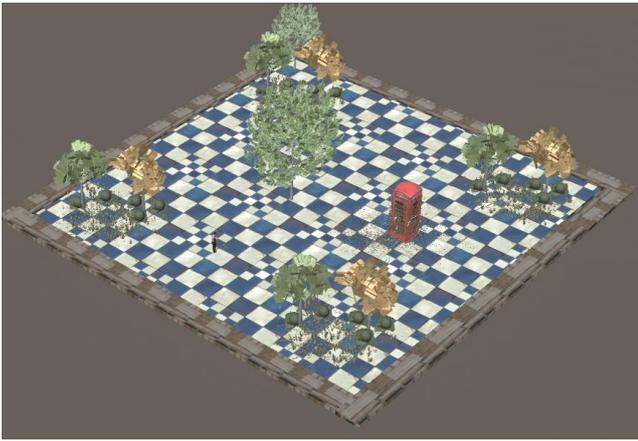
Fig. 9. Actual gameplay's 3D depiction of the real-world view from above

The player can rotate the camera and see the view from above, from different angles, as shown in the following illustrations (Figure 10):

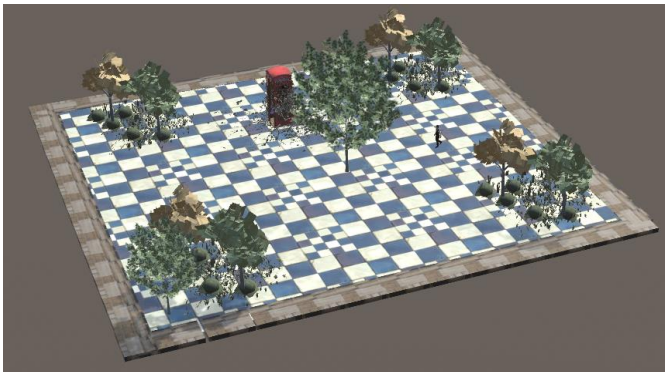
Fig. 10. Actual gameplay's 3D depiction of the real-world view from above with a player inside the scene

Furthermore, the player can change the angle of the camera and therefore the way the game is displayed, through the rotation of the mouse. The player can also enlarge or reduce the graphic display through the mouse's wheel (i.e., fluctuation of the angle of view).

The purpose of this scene is for players to understand the dilemmas and review the problems and the answers given. In particular, the element of navigating and increasing the field of vision is particularly important in this scene. Upon increasing the players' field of vision, they will be enabled to notice the three-dimensional representations (facades) of the game's levels (Figure 11).

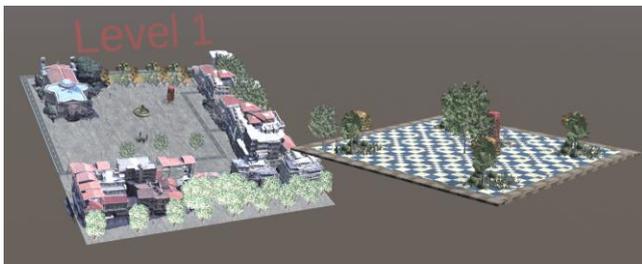
Fig. 11. Actual gameplay's 3D depiction of the real-world view where the player can preview the actual outcome of their choices i.e., how the actual world has shifted

## C. Acquiring Players' Data

At this stage, due to the small data sample and because the data are processed through SPSS, the extracted data contain less than 1000 records. While this implementation is an early record, it enables data to be extracted during the game in different types such as .csv (comma-separable files), .json, .xml, and .yml.

The data recorded focus on the player's actions and the player's movements:

a) The Actions of the players, i.e. User's Questions and Answers record the user's response and the necessary data for the identification of the specific profile i.e.:
   i. Player name = avatar name
   ii. Question Answer = choice A/B/C/... in a given dilemma
   iii. Question number = separate reference and identification code for the dilemma question (Q1, Q2)
   iv. Question description = field containing a basic description of the question and answers
   v. Timestamp = the exact date/time/second when the measurement was achieved (timestamp)

a) Users' Movements: In an attempt to further understand the player, their movements in our Gameworld are recorded to examine possible check points of interest or high value. In particular, the following are recorded.
   i. X.Axis = X-axis coordinates
   ii. Y.Axis = Y-axis coordinates
   iii. Z.Axis = Z-axis coordinates
   iv. EulerAngle (X.Rotation, Y.Rotation, Z.Rotation) = angles of view-rotation in degrees
   v. Quaternion (X.Rotation, Y.Rotation, Z.Rotation) = Unity's actual representation which is not calculated in the respective angle but via quaternion, i.e. a representation of complex numbers
   vi. TimeStamp = the exact date/time/second when the measurement was achieved (timestamp)

## VI. CONCLUSION

Our SG development focused on studying how users interact and what changes are applied when they are presented with a variety of choices (ranging from democratic/left-wing to republican/right-wing options) on particular challenges. More specifically, in the literature, evaluation typically focuses on the quality of the content, rather than the design and purpose of the application. This is an important issue since there is no universally acceptable way to measure the influence of each game on each user. For this reason, when designing any SG, there is a need to create and establish a new way of developing and evaluating, based on general standards for evaluating and testing existing games, [49], [50], [51], [52]. Our game provides a framework to study players' interaction with a digital world, both quantitively by extracting their movement/choices and qualitatively, by presenting them with overall actions and enabling them to reflect upon the impact of their decisions.

As for future steps for our game, after creating the initial templates, prefabs, and scenes, our game could expand both in terms of the cities to be visited and the questions posed. The digital city's structure should be expanded to increase the players' footprint. Moreover, questions can be customized based on questionnaire surveys, interviews as well as empirical data and suggestions from the actual players interacting with our game. Lastly, an online version of our game would make it available to a larger audience and provide users who own low-power and low-cost computer devices to participate and interact with the digital game created.

ACKNOWLEDGMENT

This work was funded in part by the research project "e-polis of the future", supported by the Hellenic Foundation of Research and Innovation (H.F.R.I.) in the context of the "1st Call for H.F.R.I. (http://www.elidek.gr) Research Projects to Support Faculty Members & Researchers and Procure High-Value Research Equipment" (Project Number: 2617). The authors would like to thank Mr. Gerasimos Kouzelis for providing the research outline of this project and Mr. Orestis Didi for his overall assistance and expertise on the topic.